# Statistical Models for Outbreak Detection of Measles in North Cotabato, Philippines

Julienne Kate N. Kintanar[1,2] and Roel F. Ceballos[1*]

[1]Mathematics and Statistics Department
University of Southeastern Philippines
Davao City, 8000 Philippines
*roel.ceballos@usep.edu.ph

[2]Research and Development Office
Central Mindanao Colleges
Kidapawan City, 9400 Philippines

## Abstract

*A measles outbreak occurs when the number of cases of measles in the population exceeds the typical level. Outbreaks that are not detected and managed early can increase mortality and morbidity and incur costs from activities responding to these events. The number of measles cases in the Province of North Cotabato, Philippines, was used in this study. Weekly reported cases of measles from January 2016 to December 2021 were provided by the Epidemiology and Surveillance Unit of the North Cotabato Provincial Health Office. Several integer-valued autoregressive (INAR) time series models were used to explore the possibility of detecting and identifying measles outbreaks in the province along with the classical ARIMA model. These models were evaluated based on goodness of fit, measles outbreak detection accuracy, and timeliness. The results of this study confirmed that INAR models have the conceptual advantage over ARIMA since the latter produces non-integer forecasts, which are not realistic for count data such as measles cases. Among the INAR models, the ZINGINAR (1) model was recommended for having a good model fit and timely and accurate detection of outbreaks. Furthermore, policymakers and decision-makers from relevant government agencies can use the ZINGINAR (1) model to improve disease surveillance and implement preventive measures against contagious diseases beforehand.*

## 1. Introduction

Measles is a highly contagious infectious disease characterized by acute respiratory viral illness. According to the World Health Organization (WHO) (2019), it is one of the world's most serious public health problems, resulting in 140,000 deaths in 2018; the majority of whom were children under the age of five. Routine measles vaccination for children below the age of five has been done to ensure that outbreaks are prevented. There have been spikes of cases and outbreaks of measles in the Philippines in recent years, indicating that the measles vaccination program needs more emphasis, specifically in far-flung and rural provinces (Ylade, 2018), including the province of North Cotabato; the reason why the authors chose this location for the study.

Routine immunizations against other contagious diseases, such as measles, have been challenging during the pandemic. As a result, the vaccination campaign for measles was put on hold for several months in 2020. The pandemic is detrimental to managing and treating other diseases, such as measles. Experts in global immunization warn of an upsurge in measles-related fatalities if vaccination efforts are not maintained (Sharmin and Rayhan, 2011).

Considering these issues, an approach to detect measles outbreaks must be employed in different localities as part of effective disease surveillance and monitoring strategies to reduce the spread of measles infection. Studies on disease surveillance models have emerged in recent years, reflecting the growing interest in the accurate and early detection of outbreaks of contagious diseases (Buckeridge, 2007). Studies around the globe have found that time series models help detect outbreaks of contagious diseases (Cliff and Haggett, 1993; Sharmin and Rayhan, 2011). In the Philippines, Paman *et al.* (2017) used Poisson integer-valued autoregressive (P-INAR) models and autoregressive integrated moving average (ARIMA) time series models in detecting measles outbreaks in Metro Manila. ARIMA's limitation includes its tendency to produce non-integer forecasts even when modeling count data like measles cases. Furthermore, P-INAR models produce integer forecasts but do not account for overdispersion and zero-inflation in the measles data. Several INAR models have been found to have robust performance when there is overdispersion; these are the new geometric first-order INAR (NGINAR) (Ristić *et al.*, 2009) and new INAR model with Bell innovations (BL-INAR) (Huang and Zhu, 2021). To address the problem of zero inflation, Ristić *et al.* (2018) proposed the use of a zero-inflated NGINAR (1) model or the ZINGINAR (1).

This study aimed to identify the measles outbreaks in North Cotabato and evaluate the performance of ARIMA and several INAR models. The conceptual advantages of P-INAR, BL-INAR, NGINAR, and ZINGINAR models in detecting measles outbreaks in North Cotabato were also be investigated.

## 2. Methodology

*2.1 Source of Data*

The dataset used for this study was provided by the Epidemiology and Surveillance Unit of North Cotabato Integrated Provincial Health Office (IPHO) upon request. The dataset contains 313 weekly measles counts from January 2016 to December 2021. Figure 1a shows the location of the province of North Cotabato, Philippines. Figure 1b shows the different municipalities covered by the province.

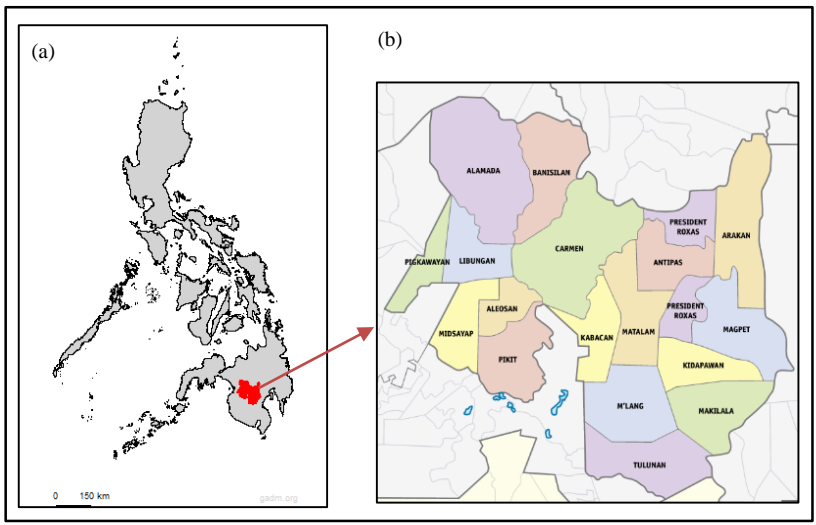

Figure 1. Map of the Philippines with North Cotabato (a) and enlarged with names of municipalities (b)

*2.2 Procedure of the Data Analysis*

2.2.1 Step 1. Identification of Outbreaks

The procedures from Rolfhamre and Ekdahl (2006) and Paman *et al.* (2017) were used to identify outbreak periods of measles in North Cotabato. The start and end of an outbreak were identified by inspecting the time series for possible outbreaks based on the weekly counts of cases of observed and recorded measles in the province. This study maximized the simplicity of defining outbreak periods based on the series mean to determine a threshold of measles cases and establish a padding measure to account for the early stages of measles. The time series mean was calculated, and two consecutive weeks with reported cases greater than the series mean was considered the start of an outbreak. The outbreak was considered over when there were fewer than the mean reported measles cases for two weeks.

Furthermore, this study defined confirmed outbreaks by reviewing official reports to determine whether these events were significant enough to warrant public health attention. Any associated reports of an outbreak were investigated for the events that were identified. Only suspected outbreaks with a corresponding official report were considered confirmed outbreaks.

2.2.2 Step 2. Application of ARIMA and INAR Models

Time series models are valuable tools for disease surveillance and outbreak detection. The ARIMA and INAR were used in this study to model the measles outbreak in the province. The ARIMA model is one of the most frequently used time series models because of its simplicity, and it has also been widely used in different studies in economics and business, engineering, technology, and disease modeling (Reis and Mandl, 2003).

The ARIMA model is a generalized version of the autoregressive moving average (ARMA) model that incorporates a differencing factor. The ARIMA model is expressed as

$$\Phi_p(B)(1-B)^d y_t = \theta_0 + \theta_q(B) a_t \qquad (1)$$

where $\Phi_p(B)$ is the operator for nonseasonal AR, the ordinary differencing factor is denoted by $(1-B)^d$, $\theta_0$ is the mean of the process, $\theta_q(B)$ is the nonseasonal MA operator, and $a_t$ is the normally distributed random shock with mean 0 and constant variance $s^2$. ARIMA modeling is an iterative process involving model identification, parameter estimation, and diagnostic checking (Montgomery *et al.*, 2016). Model identification involves determining the order of AR(p) and MA(q) model components from the stationary time series. Appropriate transformations such as differencing and log transformation are also done when the original time series exhibits nonstationarity and non-constant variance. Candidate models are then identified using the following rules:

If the ACF cut off at lag q, and the PACF should exponentially decay or show dampened sinusoidal oscillations, then it is an MA(q) process (i.e., ARIMA (0, d, q));

If the PACF cuts off at lag p, and the ACF should exponentially decay or show dampened sinusoidal oscillations, then it is an AR(p) process (i.e., ARIMA (p, d, 0));

If the ACF and PACF should exponentially decay or show dampened sinusoidal oscillations, the time series comes from an ARIMA (p, d, q) where $p, q \neq 0$.

For an illustration of the process, see Villa and Ceballos (2020). One advantage of using ARIMA is that it is computationally easy to perform. However, one of its disadvantages is that it treats discrete data as continuous, producing non-integer forecasts for modeling time series count data.

Scholars have recently proposed many models for time series count data. The most famous model was first introduced by McKenzie (1985), which is called the first-order P-INAR (1) process. The P-INAR (1) model has been helpful in epidemiology, where the number of cases of specific infectious diseases is recorded monthly or weekly to analyze and monitor acute viral infections (Silva, 2005). The P-INAR (1) model is given by

$$X_t = \alpha_1 \circ X_{t-1} + \epsilon_t \qquad (2)$$

where $\alpha$ is a scalar in (0,1) and, $\circ$ represents the binomial thinning operation; $\{\epsilon_t: t \in Z\}$ is distributed as Poisson. The characteristic of Poisson is having equal mean and variance (equi dispersion). However, in practice, many data are over-dispersed, and the variance is greater than the mean. Therefore, the Poisson-INAR is not always suitable for modeling time series count data. To address the problem of overdispersion, several authors have proposed changing the thinning operators, as summarized in Weiss (2018). Ristić *et al.* (2009) proposed using a negative binomial thinning operator and defined the INAR (1) process with geometric marginal distributions. The proposed model is often called the NGINAR (1) or the geometric first-order integer-valued autoregressive model. It is given by

$$Y_t = \alpha * Y_{t-1} + \epsilon_t \qquad (3)$$

with '∗' denote the negative binomial thinning operator. Statistical properties of the NGINAR (1) model are extensively discussed in Ristić *et al.* (2009). For more details on the statistical properties of the NGINAR (1) model, readers are directed to read their paper. Huang and Zhu (2021) proposed the BL-INAR (1) model or the Bell distributed INAR (1), which is an INAR(1) model with $\epsilon_t$ following the Bell distribution. The model was suitable for handling overdispersion in time series count data.

Aside from overdispersion, another challenge in modeling time series counts for data on infectious diseases is the presence of zero inflation or the high frequency of zeros in the dataset. To address the problem of zero inflation, Ristić *et al.* (2018) proposed the use of a zero-inflated NGINAR (1) model or the ZINGINAR (1) to deal with time series count data with more zeroes than expected. It is defined as

$$Z_t = \begin{cases} Y_t & \text{with probability } 1-\pi \\ 0 & \text{with probability } \pi \end{cases} \quad (4)$$

where $\pi \in (0,1)$ and $Y_t$ is the NGINAR (1) process. The reader is referred to Ristić *et al.* (2018) for the extensive discussion on the statistical properties of the ZINGINAR (1) model.

2.2.3 Step 3. Outbreak Detection Accuracy Measures

INAR models were evaluated for accuracy in detecting measles outbreaks. Specifically, these models were assessed in terms of the following criteria: outbreak detection accuracy (number of alarms generated, sensitivity, specificity, and false-positive rate [FPR]), and timelines (average time before detection [ATBD]).

In this study, certain thresholds and padding measures were utilized to establish criteria for identifying outbreaks and to determine the specific start and end points of these periods. Procedures from Rolfhamre and Ekdahl (2006) were applied in this study to identify outbreak periods of the actual measles data. Furthermore, the method used to determine an outbreak's start and end period was based on Paman *et al.* (2017). The series was inspected first, and periods with unexpected peaks in measles incidence were identified subjectively and marked as suspected outbreaks. Any associated reports of an outbreak were investigated for the events that were identified. Only suspected outbreaks with a corresponding report were considered confirmed outbreaks. The first of two consecutive weeks with more than three reported measles cases indicated the start of an outbreak. The outbreak was considered over when three or fewer measles cases were reported for two weeks. The series mean was used to determine the threshold of more than three reported measles cases. Alarms were generated when the series mean exceeded the threshold.

The generated outbreak detection thresholds from the fitted models were evaluated by their accuracy in detecting measles outbreaks. The number of outbreak weeks within an alarm was denoted as true positive (TP), the number of non-outbreak weeks without an alarm as true negative (TN), the number of non-outbreak weeks with an alarm as false positive (FP), and the number of outbreak weeks without an alarm as false negative (FN). Sensitivity, specificity, false positive rates, and positive and negative predictive values were used to measure the accuracy of the models in detecting measles outbreaks. Furthermore, the timelines of detection were measured in terms of ATBD.

## 2.3 Framework for Modeling Implementation

The flowchart provided in Figure 2 summarizes the implementation of different models for the measles.

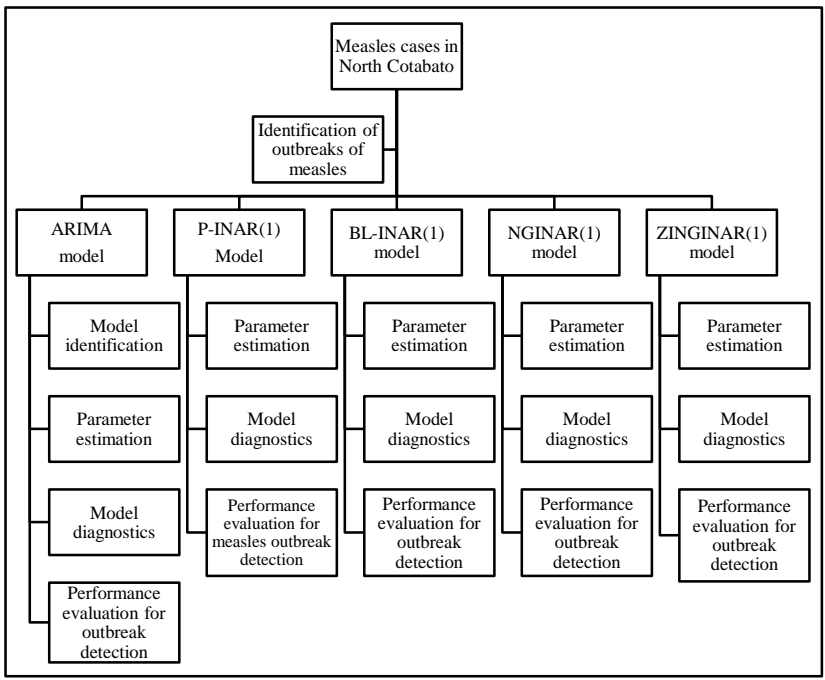

Figure 2. Flowchart for modeling measles outbreak in North Cotabato

Weekly measles cases in the province were used in the analysis. Outbreaks were identified based on the procedures described in the previous section. Different INAR models, together with ARIMA, were developed for the purpose of detecting measles outbreaks in the province. Models were assessed in terms of model fit and performance to detect measles outbreaks, and these are included in the discussions of results.

Datasets were split into two: the first half, which contained the first outbreak period, was used for training or estimating parameters, and the second set, which contained the second outbreak period, was utilized for performance evaluation of the models. The parameter estimates were generated using the method of maximum likelihood. All calculations were done using the R programming language (v4.1.3) (R Core Team, 2022).

## 3. Results and Discussion

### 3.1 Identification of Outbreaks

Two identified measles outbreaks occur in North Cotabato, as shown in Figure 3. The first outbreak started on February 4, 2018, and ended on June 30, 2019. The second outbreak started on August 25, 2019, and ended on March 29, 2020. The start of the first outbreak had nine cases and was determined by the first of two consecutive weeks with more than three cases. The end of the first outbreak had one case, which was determined by the last of two consecutive weeks with three or fewer reported measles cases. The second outbreak started with seven cases and ended with one case by implementing the same padding measures as the first.

The first outbreak was confirmed through reports issued by the Department of Health (DOH) (Fernandez, 2019). In 2018, DOH-12 recorded 328 measles cases in the provinces of North Cotabato, Kidapawan City, South Cotabato, and other cities in Cotabato. To address the outbreak, the IPHO personnel conducted a house-to-house vaccination in the barangays to ensure that children nine months to five years old are given vaccines (Fernandez, 2019). The second outbreak was also identified as a confirmed outbreak. A supporting report showed that North Cotabato recorded 211 measles cases with 11 fatalities. As an intervention, they had intensified the anti-measles campaign due to the higher number of cases recorded from January 2020 to October 2020 (Estabillo, 2020). It is important to note that the second outbreak coincides with the early COVID-19 outbreak in the Philippines; during which, most communities had mobility restrictions. The community lockdowns and limited mobility affected the conduct of immunization against measles. In the face of a pandemic, it became crucial to recognize the urgency of designing and implementing comprehensive programs prioritizing the uninterrupted immunization of children against communicable diseases. The continuation of immunization efforts was of utmost importance to safeguard the health and well-being of the young population.

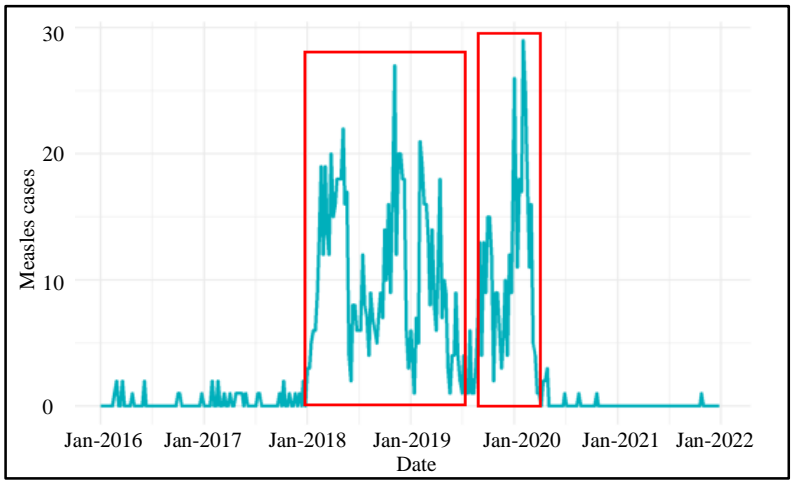

Figure 3. Time series plot of measles data in North Cotabato

*3.2 Parameter Estimates of the ARIMA Model*

The time series plot of measles cases in North Cotabato showed an irregular pattern, suggesting that the series was not stationary (Figure 4). Log transformation and differencing were done on the measles data to address the issue of nonstationary and non-constant variance.

The result of the ADF test in Figure 4a revealed that the transformation works, and the measles data was already stationary. The next step was to determine the order of AR and MA terms, which might help correct the remaining autocorrelations and aid in determining the candidate models. The plot of the ACF and PACF presented in Figure 4b was used to determine the candidate models using the guidelines proposed by Montgomery *et al.* (2016).

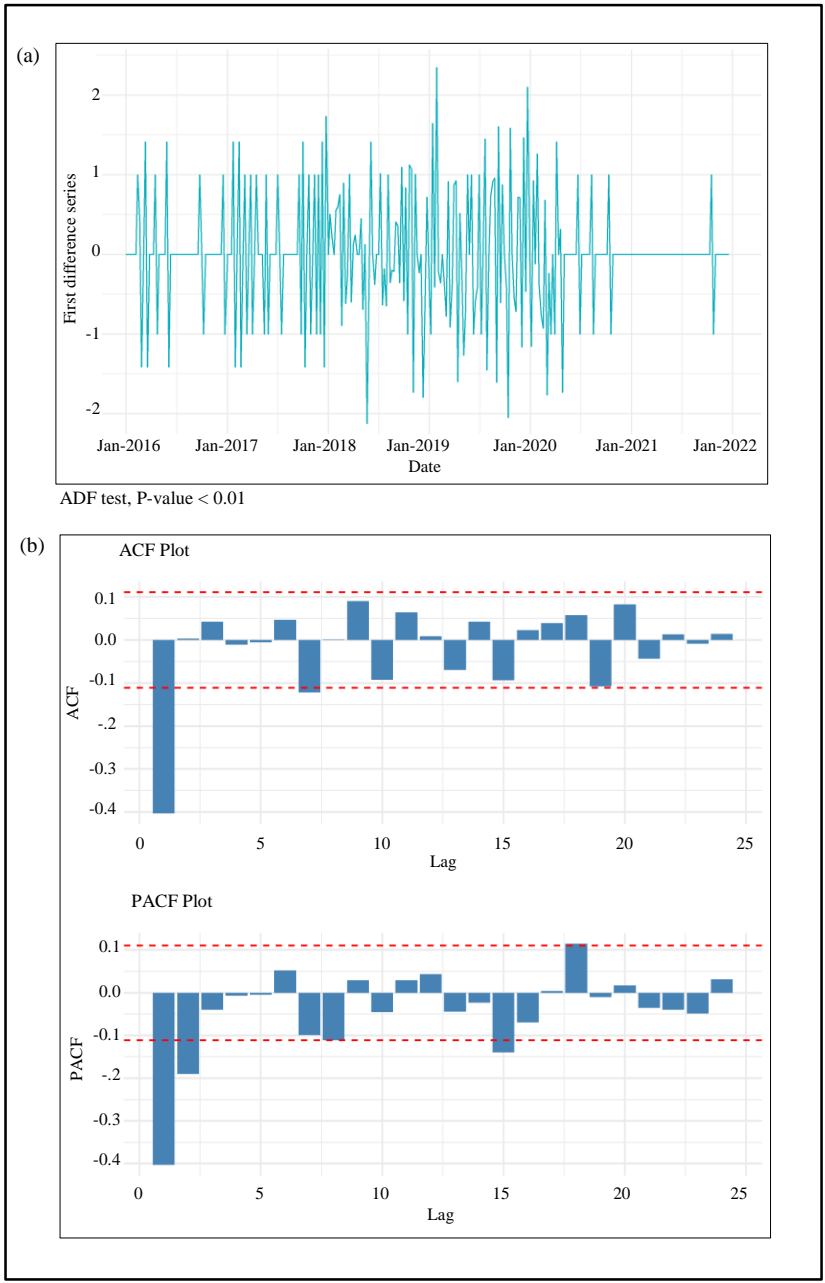

Figure 4. Plot of transformed series (a); ACF and PACF plot of the stationary series (b)

Looking at Figure 4b's ACF plot, autocorrelation appeared to cut off after lag 1. When examining the PACF, it had an oscillating pattern and tailed off at 0; ARIMA (0,1,1), or possibly ARIMA (1,1,1) was therefore considered. Another interpretation is that the PACF plot tailed off after lag 1 or 2 and ACF cut off at lag 1, indicating an ARIMA (2,1,1) model. Also, the PACF tailed off, suggesting ARIMA (1,1,0) or ARIMA (2,1,0) model.

To identify the best ARIMA model among the tentative models for the measles data in North Cotabato, the Akaike Information Criterion (AIC) was used. Table 1 presents the tentative ARIMA models for the weekly measles cases in the province and their respective AIC values. The best model in the list was ARIMA (0, 1, 1) since it had the smallest AIC value. In Table 2, the estimates were significantly different from zero, and in Table 3, the Ljung-Box test confirmed that the model may be useful (P-value = 0.4658). However, Figures 5a and 5b show that the residual plots had a funneling pattern, suggesting that the model fit might be problematic, which was expected since ARIMA is not very suitable for modeling counts of cases for infectious diseases.

Table 1. AIC values of the candidate models

| Model | AIC Value |
|---|---|
| ARIMA (0,1,1) | 1622 |
| ARIMA (1,1,1) | 1624 |
| ARIMA (1,1,0) | 1625 |
| ARIMA (2,1,0) | 1624 |
| ARIMA (2,1,1) | 1626 |

Table 2. Test on the model estimates

|  | Estimates | Std. error | z-value | P-value |
|---|---|---|---|---|
| MA (1) | -0.3883 | 0.0509 | -7.63 | < 0.01 |

Table 3. Ljung-Box test of the residuals

| Test statistic | Degrees of freedom | P-value |
|---|---|---|
| -0.3883 | 0.0509 | 0.04658 |

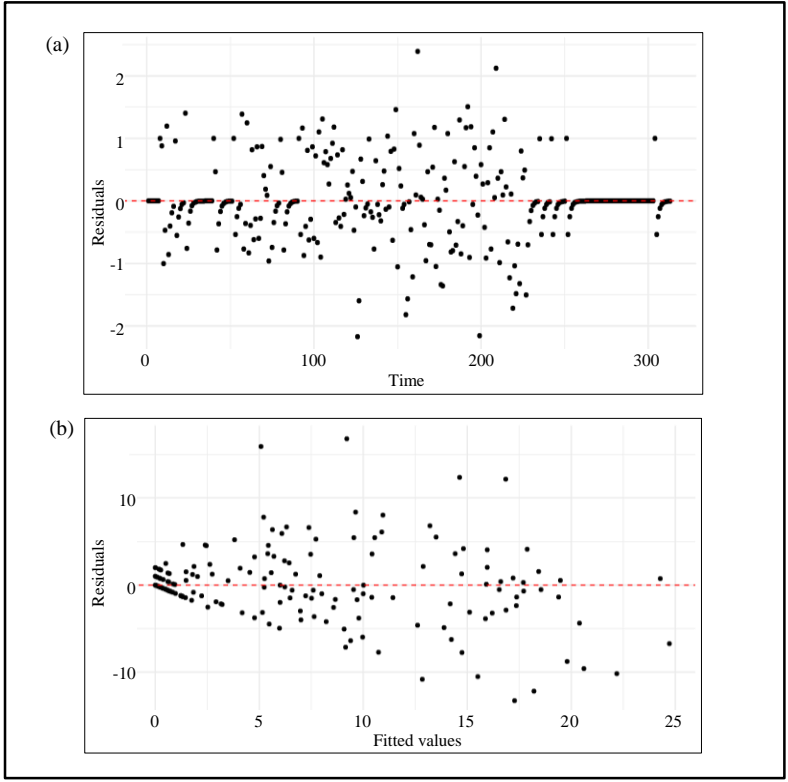

Figure 5. Plot of residuals vs time (a) and plot of residuals vs. predicted values (b)

## 3.3 Parameter Estimates of the INAR Models

Overdispersion is defined as greater variability in observations than expected from a given model. A Poisson distribution is commonly assumed for INAR models, comparable to the normal distribution for real-valued time series McKenzie (1985) and Al-Osh and Alzaid (1987). The Poisson index of dispersion was calculated, showing that the series had a sample mean of 4 per week and exhibited a poisson index of dispersion $I = 9.9$. This index indicated that the measles series was likely from an over dispersed distribution concerning Poisson distribution. Hence, models that could handle overdispersion were more appropriate in this case.

The ZINGINAR (1) model had a parameter estimate for $\pi$ equal to 0.2532, suggesting that the measles series exhibited an excess in zeroes of about 25% compared with the NGINAR (1) (Table 4). Furthermore, there was an average

143

of 2 ($\hat{\mu} = 1.741$) measles reported cases per week in North Cotabato and around 78.58% ($\hat{\alpha} = 0.7858$) of the variation in measles counts from one week to the next can be explained by the count in the previous week. The P-value of the Vuong test ($p = 0.0281$) strongly suggested that ZINGINAR (1) model was more appropriate than NGINAR (1) due to the presence of zero-inflation. Hence, ZINGINAR (1) was considered an appropriate model.

Table 4. Parameter estimates of the models

| P-INAR (1) | | BL-INAR (1) | | NGINAR (1) | | ZINGINAR (1) | |
|---|---|---|---|---|---|---|---|
| $\alpha$ | 0.7122 | $\alpha$ | 0.6571 | $\alpha$ | 0.7315 | $\alpha$ | 0.7858 |
| $\lambda$ | 0.8914 | $\theta$ | 0.3257 | $\mu$ | 1.632 | $\mu$ | 1.741 |
| | | | | | | $\pi$ | 0.2532 |

From the estimates of presented in Table 4, the ZINGINAR (1) model can be written as

$$\hat{Z}_t = \begin{cases} \hat{Y}_t, & \text{with probabiliy } 0.7468 \\ 0, & \text{with probability } 0.2532 \end{cases} \quad (5)$$

where $\hat{Y}_t = 1.741 + 0.7858 * Y_{t-1}$ and $\hat{Y}_t$ is a NGINAR (1) process.

In Table 5, ARIMA (0,1,1) had the lowest MAE, RMSE, and AIC among the competing models. However, ARIMA (0,1,1) generated non-integer forecasts that did not reflect the natural dynamics of measles cases. In this regard, ARIMA (0,1,1) was not the most appropriate model. NGINAR (1) and ZINGINAR (1) had considerably lower AICs than the P-INAR (1) and BL-INAR (1) models. Furthermore, the ZINGINAR (1) had lower MAE, RMSE, and AIC than the NGINAR (1) model. Hence, in this case, ZINGINAR (1) was preferred.

Table 5. MAE, RMSE, and AIC

| Measures | P-INAR (1) | BL-INAR (1) | NGINAR (1) | ZINGINAR (1) | ARIMA (0,1,1) |
|---|---|---|---|---|---|
| MAE | 1.6955 | 1.5281 | 1.7551 | 1.4784 | 1.4172 |
| RMSE | 4.018 | 3.954 | 4.585 | 4.411 | 3.629 |
| AIC | 1874 | 1775 | 1358 | 1274 | 1622 |

*3.4 Performance Evaluation of Outbreak Detection Models*

Using the models and parameter estimates, looking into the forecasts during the second outbreak was also interesting. The second outbreak datasets were not included in the model building or parameter estimation. The MAE and RMSE of these models are summarized in Table 6. For these comparisons, the ARIMA (0,1,1) and ZINGINAR (1) models had considerably lower MAE and RMSE values compared with the rest of the models. Hence, the ZINGINAR (1) model was preferred in this case since it produced non-negative forecasts, reflecting the true nature of measles count data. Furthermore, ZINGINAR (1), having the lowest MAE and RMSE values, proved that it was better to address overdispersion and zero inflation in modeling measles count data.

Table 6. MAE and RMSE

| Measures | P-INAR (1) | BL-INAR (1) | NGINAR (1) | ZINGINAR (1) | ARIMA (0,1,1) |
|---|---|---|---|---|---|
| MAE | 3.581 | 4.152 | 3.312 | 3.415 | 3.114 |
| RMSE | 7.172 | 6.025 | 5.706 | 4.971 | 4.681 |

Table 7 presents the different outbreak detection measures for measles data using the different models. A general characteristic of the INAR models is that they generate more alarms than the ARIMA model. More alarms relate to higher sensitivity and higher false positive rates. Regarding timeliness, the ARIMA (0,1,1) model did not detect the outbreak during its first week, and it took around five weeks or more than a month before an outbreak was detected by this model, which was undesirable. Furthermore, all INAR models had an average time before detection of 0 week, which suggested that these models effectively detected an outbreak right at the start, which was desirable for appropriate action to mitigate the effects of a measles outbreak.

Table 7. Outbreak detection measures

| Measures | P-INAR (1) | BL-INAR (1) | NGINAR (1) | ZINGINAR (1) | ARIMA (0,1,1) |
|---|---|---|---|---|---|
| Alarms | 49 | 42 | 31 | 28 | 10 |
| Sensitivity | 70.00 | 62.52 | 20.18 | 15.47 | 9.85 |
| Specificity | 93.84 | 95.62 | 97.15 | 98.96 | 99.25 |
| FPR | 6.16 | 4.38 | 2.85 | 1.04 | 0.75 |
| Detection on week1 | Yes | Yes | Yes | Yes | No |
| ATBD (weeks) | 0 | 0 | 0 | 0 | 5 |
| PPV | 49.62 | 51.24 | 58.26 | 69.28 | 92.85 |
| NPV | 92.54 | 91.25 | 87.25 | 88.81 | 87.25 |

## 4. Conclusion and Recommendation

This study demonstrated the applicability of the ARIMA and INAR models in detecting measles outbreaks in North Cotabato. INAR models had the conceptual advantage of producing an integer forecast, which reflected the natural dynamics of measle data. All INAR models showed great potential in timely detecting measles outbreaks in the province. While BL-INAR (1), NGINAR (1), and ZINGINAR (1) can model overdispersion in measles data, only the ZINGINAR (1) model can model excess zeroes (zero inflation). This is the conceptual advantage of the ZINGINAR (1) model compared with other INAR models. Furthermore, the ZINGINAR (1) model had generally lower AIC, MAE, and RMSE than other INAR models and performed better in forecasting accuracy. Thus, ZINGINAR (1) was considered an optimal model for measles outbreak detection since it generated non-negative integer values and is appropriate for count time series data, especially for over-dispersed data with zero inflation.

This study showed the applicability and performance of INAR models in the outbreak detection of measles in North Cotabato. Based on these results, concerned agencies in the prevention and control of measles in the province can use the ZINGINAR (1) model to improve disease surveillance and monitoring to effectively implement preventive measures against measles outbreaks.

Future studies may explore the applicability of the INAR models in detecting outbreaks of other infectious diseases. They may also explore another modeling framework for comparative purposes, such as the Time Series Susceptible, Infectious, and Recovered (TSIR).

## 5. Acknowledgment

The researchers would like to acknowledge the Integrated Provincial Health Office – Provincial Epidemiology and Surveillance Unit (IPHO-PESU) of North Cotabato for providing the data used in this study.